\documentclass[twocolumn]{jetpl}

\usepackage{amsmath}
\usepackage{amssymb}
\usepackage{amsfonts}
\usepackage{graphicx}
\usepackage{color}
\usepackage{bm}
\usepackage{cite}

\renewcommand{\Re}{\hbox{\rm Re}\,}
\renewcommand{\Im}{\hbox{\rm Im}\,}

\begin{document}

\title{Out-of-Equilibrium Admittance of Single Electron Box Under Strong Coulomb Blockade}

\rtitle{Out-of-Equilibrium Admittance of Single Electron Box Under Strong Coulomb Blockade}

\sodtitle{Out-of-Equilibrium Admittance of Single Electron Box Under Strong Coulomb Blockade}

\author{Ya.\,I.\,Rodionov and I.\,S.\,Burmistrov
\thanks{e-mail: yaroslav.rodionov@gmail.com burmi@itp.ac.ru.}}

\dates{\today}{*}

\address{
L.D. Landau Institute for Theoretical Physics RAS, 119334 Moscow, Russia}

\abstract{We study admittance and energy dissipation in an
out-of-equlibrium single electron box. The system consists of a
small metallic island coupled to a massive reservoir via single
tunneling junction. The potential of electrons in the island is
controlled by an additional gate electrode. The energy dissipation
is caused by an AC gate voltage. The case of a strong Coulomb
blockade is considered. We focus on the regime when electron
coherence can be neglected but quantum fluctuations of charge are
strong due to Coulomb interaction.  We obtain the admittance under
the specified conditions. It turns out that the energy dissipation
rate can be expressed via charge relaxation resistance and
renormalized gate capacitance even out of equilibrium. We suggest
the admittance as a tool for a measurement of the bosonic
distribution corresponding collective excitations in the system.}


\maketitle

It is well-known that the phenomenon of Coulomb blockade is an
excellent tool for observation of interaction effects in single
electron devices~\cite{zaikin,ZPhys,grabert,aleiner,Glazman}.
Recently, due to the progress in the field of thermoelectricity the
Coulomb blockade under out-of-equilibrium conditions has come into
the focus of the theoretical~\cite{BaskoKravtsov,Bagrets,AltlandEgger,Nazarov,Pekola,rodbur2} and experimental
research~\cite{Giazotto_review,ScheibnerNew,Hoffmann}. The simplest mesoscopic system displaying Coulomb blockade is a
single electron box (SEB). The properties of such a system are
essentially affected by electron coherence and interaction.

\begin{figure}[b]
  \centerline{\includegraphics[width=70mm]{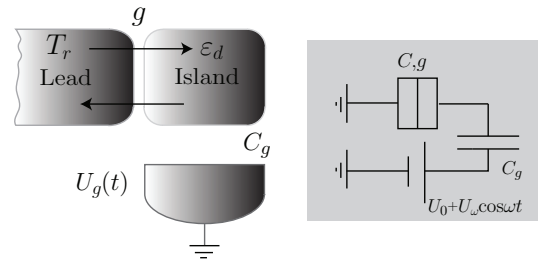}}
  \caption{Fig.~\ref{figure1}.
    %Measurement of the resistance $R_q$.
    The set-up. A SEB is subjected to a constant gate voltage\ $U_0$.
    The current through the tunneling junction is caused by a weak
    AC voltage\ $U_g(t)$.
          }\label{figure1}
\end{figure}

The set-up is as follows (see Fig.\ref{figure1}). Metallic island is
coupled to an equilibrium electron reservoir (temperature $T_r$) via
tunneling junction. The island is also coupled capacitively to the
gate electrode. The potential of the island is controlled by the
voltage\ $U_g$\ of the gate electrode. The distribution function of
electrons in the reservoir is assumed to be equilibrium (Fermi
distribution) while the one inside the island is arbitrary.

The physics of the system is governed by the Thouless energy of an
island\ $E_{\rm Th}$, its charging energy\ $E_c$, and the mean single-particle level
spacing\ $\delta$. Throughout the paper the Thouless energy is
considered to be the largest scale in the problem. This allows us to
treat the metallic island as a zero dimensional object with
vanishing internal resistance. The characteristic energy ($\varepsilon_d$)
of electrons inside the island obeys the condition $\delta\ll
\varepsilon_d\ll E_c,E_{\rm Th}$. This implies that characteristic energy is
high enough to render the system incoherent and low enough to keep
it strongly correlated due to Coulomb interaction~\cite{BEBS}. The dimensionless
conductance of a tunneling junction is small $g\ll1$.

A single electron box does not allow for conductance measurements since there is no
DC-transport. This way an essential dynamic characteristic becomes
the set-up admittance, which is a current response to an AC-gate
voltage $U_g(t)=U_0+U_\Omega \cos\Omega t$.

Paper~\cite{buttiker0} sparked both theoretical and experimental
attention to the admittance
%dynamic response functions
of such a
set-up~\cite{buttiker3,buttiker2,imry,Park,gabelli,delsing}.
As it is well-known, the real part of admittance determines energy
dissipation in an electric circuit. Classically, the average energy
dissipation rate of a single electron box is given as follows
\begin{gather}
 \label{admittance0}
\mathcal{W}_\Omega=\Omega^2C^2_g R\frac{|U_\Omega|^2}{2},\quad
R=\frac{h}{e^2 g},\quad
 \hbar \Omega\ll gE_c,
\end{gather}
where $C_g$ denotes the gate capacitance, $e$ - the electron charge,
and $h=2\pi \hbar$ - the Planck constant. Expression
\eqref{admittance0} presents us with a natural way of extracting the
resistance of a system from its dissipation rate.

%%%%%%%%%%%%%%%%%%%%%%
%%%%%%%%%%%%%%%%%%%%%%%%

Electron coherence and interaction change the classical result for
the dissipation. The low-temperature ($T\ll\delta$) coherent regime
was pioneered in Ref.~\cite{buttiker0}. It was shown that the energy
dissipation rate $\mathcal{W}_\Omega$ can be factorized in
accordance with its classical appearance \eqref{admittance0} but the
definition of physical quantities comprising it becomes different.
Geometrical capacitance\ $C_g$\ should be substituted by a new
observable: mesoscopic capacitance\ $C_\mu$. This leads to the
establishment of another observable: {\it charge relaxation}
resistance\ $\mathcal{R}_q$ such that \ $R\rightarrow \mathcal{R}_q$
in Eq.~\eqref{admittance0}. Charge relaxation resistance of a
coherent system differs drastically from its classical counterpart.
In particular, the charge relaxation resistance of a single channel
junction was predicted to be independent of its transmission and
equal to $h/(2e^2)$ at zero temperature~\cite{buttiker0}. However
Coulomb interaction in Ref.~\cite{buttiker0} and subsequent
works~\cite{buttiker2} was accounted for on the level of classical
equations of motion only. Recently the result for quantization of
the charge relaxation resistance in SEB at $T\ll \delta$ has been
rigorously derived~\cite{lehur}. The admittance in this low
temperature regime was investigated experimentally by Gabelli et
al.~\cite{gabelli}.

The knowledge of the charge relaxation resistance has been extended
to a SEB at the transient temperatures when thermal fluctuations
smear out electron coherence but electron-electron interaction is
strong. The expression for the energy dissipation at this transient
temperatures keeps its classical appearance if one substitutes the
renormalized gate capacitance $\mathcal{C}_g$ and the charge
relaxation resistance $\mathcal{R}_q$ for $C_g$ and $R$
respectively~\cite{rodbur}. Unlike the latter, $\mathcal{C}_g$ and
$\mathcal{R}_q$ have strong temperature and gate voltage $U_g$
dependance.

The recent experiment by Persson et al~\cite{delsing} explored the
energy dissipation rate at these transient temperatures.  The
admittance of SEB was measured at fixed frequency as a function of
 pumping amplitude $U_\Omega$ and the DC part of gate voltage $U_0$
 in a wide range.
  The theoretical analysis of the data in Ref.~\cite{delsing}
  was carried out under assumption of linear response to the AC gate
  voltage: the electrons inside the island were assumed to be in the equilibrium
  with the reservoir. However it has not been verified experimentally.
  It is natural to expect that this assumption is violated for the
 set of data with high values of the amplitude $U_\Omega$.

Motivated by the experiment~\cite{delsing} we study the admittance
of a single electron box under the out-of-equilibrium conditions. We
consider the linear response of a SEB with \emph{arbitrary} electron
distribution function in the island to the AC gate voltage.

%%%%%%%%%%%%%%%%%%%%%%%%%%%%%%%%%%%%%%%%%%%%%%%%%%%%%%%%%%%%%%%%%%%%%
%%%%%%%%%%%%%%%%%%%%%%%%%%%%%%%%%%%%%%%%%%%%%%%%%%%%%%%%%%%%%%%%%%%%%
%%%%%%%%%%%%%%%%%%%%%%%%%%%%%%%%%%%%%%%%%%%%%%%%%%%%%%%%%%%%%%%%%%%%%
A single electron box
 is described by the Hamiltonian
\begin{gather}
   \label{ham1}
      H=H_0+H_c+H_t ,
\end{gather}
where
\begin{gather}
   \label{ham2}
    H_0=\sum_{k,i}\varepsilon_{k}a^{\dagger}_{k}a_{k}+
    \sum_\alpha\varepsilon^{(d)}_\alpha d^\dagger_{\alpha}d_{\alpha}
\end{gather}
describes free electrons in the lead and the island,\ $H_c$\
describes Coulomb interaction of carriers in the island, and\ $H_t$\
describes the tunneling. Here operators $a^{\dag}_{k}$
($d^\dag_{\alpha}$)\ create a carrier in the lead (island). Then the
tunneling Hamiltonian is
\begin{gather}
   \label{ham-tun}
    H_{t}=X+X^\dagger,\ \ X=\sum_{k,\alpha}t_{k\alpha} a_{k}^{\dagger}
    d_{\alpha}.
\end{gather}
The charging Hamiltonian of electrons in the box is taken in the
capacitive form:
\begin{gather}
   \label{ham3}
      H_c=E_c\big(\hat{n}_d-q\big)^2.
\end{gather}
Here $E_c= e^2/(2C)$\ denotes the charging energy, $q=C_g U_g/e$\ the gate charge, and $\hat{n}_d$\ is an operator of a
particle number in the island $\hat{n}_d=\sum_{\alpha}d^\dagger_{\alpha}d_{\alpha}$.
%:
%\begin{gather}
%   \label{number}
%    \begin{split}
%    \hat{n}_d=\sum_{\alpha}d^\dagger_{\alpha}d_{\alpha}.
%    \end{split}
%\end{gather}
To characterize the tunneling it is convenient to introduce the
Hermitean matrix:
\begin{gather}
   \check{{g}}_{\alpha\alpha^\prime}=(2\pi)^2
   \left[\delta(\varepsilon^{(d)}_{\alpha})\delta(\varepsilon^{(d)}_{\alpha^\prime})
   \right]^{1/2}\sum_k t^\dagger_{\alpha k}
   \delta(\varepsilon_{k})t_{k\alpha^\prime}.\label{gaa}
\end{gather}
The energies $\varepsilon_k,\varepsilon^{(d)}_\alpha$ are accounted
from the Fermi level, and the delta-functions should be smoothed on
the scale $\delta E$, such that $\delta \ll\delta E\ll T_r,\varepsilon_d$.
The classical dimensionless conductance (in units $e^2/h$) of the
junction between a reservoir and the island can be expressed as
follows
%
%\begin{gather}
% \label{conductance-def1}
 $g=\sum_\alpha \check{g}_{\alpha\alpha}$.
%\end{gather}
%
Therefore, each non-zero eigenvalue of $\check g$ corresponds to the
transmittance of some `transport' channel between a reservoir and
the island~\cite{landauer}. The effective dimensionless conductance ($g_{\rm ch}$) of a
`transport' channel and their effective number ($N_{\rm ch}$) are given by
\begin{equation}
 g_{\rm ch}=
\frac{\sum\limits_{\alpha\alpha^\prime} \check
g_{\alpha\alpha^\prime} \check g_{\alpha^\prime
\alpha}}{\sum\limits_\alpha \check{g}_{\alpha\alpha}} ,\quad N_{\rm ch} = \frac{\left ( \sum\limits_\alpha
\check{g}_{\alpha\alpha} \right )^2}
{\sum\limits_{\alpha\alpha^\prime} \check g_{\alpha\alpha^\prime}
\check g_{\alpha^\prime \alpha}} .
\end{equation}
The dimensionless conductance becomes
%\begin{equation}
$g = g_{\rm ch}N_{\rm ch}$.
%\end{equation}
%
In what follows we will always assume
\begin{gather}
\label{aes-condition2} g_{\rm ch}\ll 1, \qquad N_{\rm ch}\gg 1,\quad
g\ll1.
\end{gather}
Throughout the paper we keep the units such that $\hbar=e=k_B=1$ except for the final results.

%%%%%%%%%%%%%%%%%%%%%%%%%%%%%%%%%%%%%%%%%%%%%%%%%%%%%%%%%%%%%%%%%%%%
%%%%%%%%%%%%%%%%%%%%%%%%%%%%%%%%%%%%%%%%%%%%%%%%%%%%%%%%%%%%%%%%%%%%
%%%%%%%%%%%%%%%%%%%%%%%%%%%%%%%%%%%%%%%%%%%%%%%%%%%%%%%%%%%%%%%%%%%%
In the presence of time dependent gate voltage the gate charge $q$
in Eq.~\eqref{ham3} is changed as $q\rightarrow C_gU_g(t)/e$. The
gate voltage is coupled to the operator of particle number inside
the island only. Therefore the admittance of the system (the
response of the charge in the island to AC part of the gate voltage)
is determined by autocorrelation function of fluctuating particle
number: $i \theta(t) \langle[\hat n_d(t),\hat n_d(0)]\rangle$, where
$\theta(t)$ is Heaviside step-function. Due to the presence of
strong Coulomb interaction the behavior of the autocorrelation
function is non-trivial. It corresponds to collective bosonic modes
similar to the case of Fermi liquid where the density-density
correlator is governed by the electron-hole
excitations~\cite{AGD,Mahan,Kamenev_rev}. The latter determines the
behavior of the autocorrelation function in the absence of the
Coulomb interaction. In an out-of-equilibrium regime we generally
expect the collective mode distribution to be different from the
distribution of the electron-hole excitations. As shown in
Ref.~\cite{rodbur2}, the collective mode distribution coincides with
the one for the electron-hole excitations even out of equilibrium:
\begin{equation}
  \label{e-h-distr}
   B_\omega(\tau)=\frac{\int\big[1-F^d_\varepsilon(\tau)
        F^r_{\varepsilon-\omega}(\tau)\big]d\varepsilon}
        {\int
   \big[F^d_\varepsilon(\tau)-F^r_{\varepsilon-\omega}(\tau)\big]
   d\varepsilon}.
\end{equation}
Here function\ $F^{d,r}_\varepsilon(\tau)$\ is given in terms of the
Wigner transform $f^{d,r}_\varepsilon(\tau)$ of the  electron
distribution function\ $f^{d,r}(t,t^\prime)$ inside the
island/reservoir:
$F^{d,r}_\varepsilon(\tau)=1-2f^{d,r}_\varepsilon(\tau)$, where a
slow time $\tau=(t+t^\prime)/2$. In the equilibrium
$F^{d,r}_\varepsilon=\tanh(\varepsilon/2T_r)$ and
$B_\omega=\coth(\omega/2T_r)$.

\begin{figure}[t]
   \centerline{\includegraphics[width=65mm]{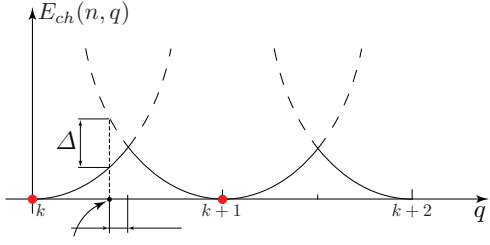}}
       \caption{Fig.~\ref{figure2}.
   Charging energy\ $E_{ch}=E_c(n-q)^2$\ as a function of gate charge $q$.
          } \label{figure2}
\end{figure}

\emph{Results.}  -- We focus on the most interesting case of Coulomb
peak: the vicinity of a degeneracy point\ $q=k+1/2$\ where $k$ is an
integer. In this parametric regime the transport is dominated by the
two closest charging states~\cite{matveev} (see Fig.~\ref{figure2})
which in the case of $g=0$ are separated by the Coulomb gap
$\Delta=2E_c(k+1/2-q)$. Due to the presence of the tunneling (finite
$g$) all the  observables, e.g., $\Delta$, undergo strong
renormalization near the Coulomb peak.

%%%%%%%%%%%%%%%%%%%%%%%%%%%%%%%%%%%%%%%
%%%%%%%%%%%%%%%%%%%%%%%%%%%%%%%%%%%%%%

 For not very high frequencies $\Omega \ll\max\{|\bar{\Delta}|,T_r,\varepsilon_d\}$
 we obtained the following expression for admittance of the SEB
\begin{gather}
  \label{admittance_rate}
  \mathcal{G}_\Omega=\frac{C_g}{C}\frac{Z^4\bar{g}}{4\pi}
  \frac{\bar{\Delta}\partial_{\bar{\Delta}}B_{-\bar{\Delta}}}{B_{-\bar{\Delta}}}
  \frac{i\Omega}{-i\Omega-\frac{\bar{g}{\bar\Delta}B_{-{\bar\Delta}}}{2\pi}}.
\end{gather}
Here the scaling parameter $Z$ is defined as
\begin{gather}
    \label{green1_1}
   Z(\lambda)=\Bigl (1+\frac{g}{2\pi^2}\lambda \Bigr )^{-1/2}, \quad
   \lambda=\int\frac{{B}_{\omega}}{2\omega}d\omega ,
\end{gather}
and $\bar{g},\bar{\Delta}$ are renormalized tunneling conductance
and Coulomb gap respectively:
\begin{gather}
  \label{green1_2}
\bar{g}=g Z^2(\lambda),\qquad \bar{\Delta}=\Delta Z^2(\lambda).
\end{gather}
The integral in Eq.~\eqref{green1_1} runs over frequencies $E_c\gg
|\omega|\gg \omega_0 = \max\{T_r,\varepsilon_d,|\bar{\Delta}|\}$.
The energy scale $\omega_0$ determines the natural scale at which
the RG procedure has to be stopped~\cite{rodbur2}. Within
logarithmic accuracy we find $\lambda=\ln{E_c/\omega_0}$.

%%%%%%%%%%%%%%%%%%%%%%%%
\begin{figure}[b]
\centerline{\includegraphics[width=80mm]{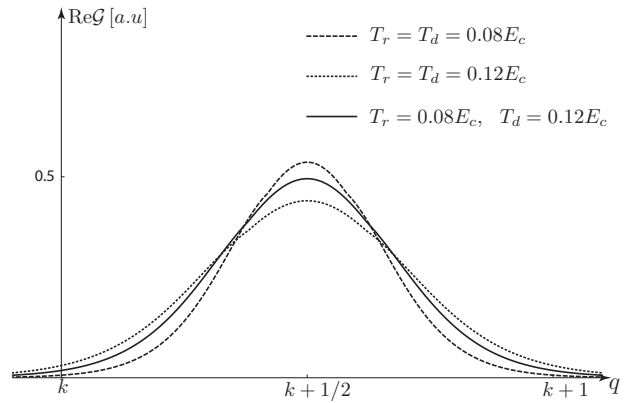}}
  \caption{Fig.~\protect\ref{figureAdm}
    The real part of admittance of the SEB at fixed $\Omega$
as a function of $q$. We use $g=0.5$, $\Omega= 0.02 E_c$ and
$C_g/C=0.24$. See text.
             }\label{figureAdm}
\end{figure}
%%%%%%%%%%%%%%%%%%%%%%%

%%%%%%!!!!!!!!!!!!!!
 We stress that our result~\eqref{admittance_rate} is valid for
 an arbitrary electron distribution. To make
 predictions more  concrete  we consider the case of quasi-equilibrium
 $F_\varepsilon^d=\tanh\varepsilon/2T_d,\ T_d>T_r$ as an example.
 This regime is typical for a SEB with the metallic island.
 It is achieved when the energy relaxation rate due to
 electron-electron interaction in the island
 $1/\tau_{ee} \gg g\delta$ (see e.g.,~\cite{rodbur2}).

%%%%%%!!!!!!!!!!!!!
The real part of admittance~\eqref{admittance_rate} at fixed
$\Omega$ as a function of $q$ is shown in Fig.~\ref{figureAdm} for
the out-of-equilibrium regime with $T_d>T_r$. At fixed $C_g$, $C$
and $g$ the height of the maximum  is controled by the effective
temperature of electron-hole excitations $T_{\rm eh} =
\lim_{\bar\Delta\to 0} (\bar\Delta/2)
B_{\bar\Delta}$~\cite{Chtchelkatchev}. As it was shown,
$T_r\leqslant T_{\rm eh} \leqslant T_d$ and $T_{\rm eh} \approx T_d
\ln 2$ for $T_d\gg T_r$ \cite{rodbur2}. Therefore,
out-of-equilibrium admittance is confined within the boundaries
$\Re{\cal G}_{\Omega,T_d}<\Re{\cal G}_\Omega<\Re{\cal
G}_{\Omega,T_r}$, where ${\cal G}_{\Omega,T_d}({\cal
G}_{\Omega,T_r})$ are equilibrium admittances at temperatures
$T_d(T_r)$.

\begin{figure}[t]
\centerline{\includegraphics[width=80mm]{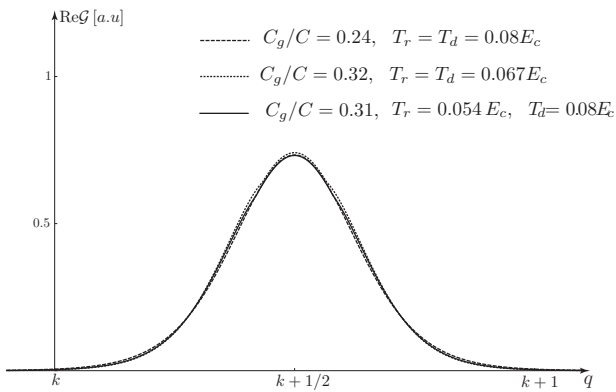}}
 \caption{Fig.~\protect\ref{figureDelsing}
   The dissipative part of admittance of the SEB at fixed $\Omega$
as a function of $q$. Three curves corresponding to three different
formulae are presented. Dashed line corresponds to
Eq.~\eqref{admittance_rate} with $Z=1$, $\bar{\Delta}=\Delta$,
$\bar{g}=g$ and $B_{\bar\Delta} = \coth \Delta/2T_r$. Dotted line is
plotted according to Eq.~\eqref{admittance_rate} with
$B_{\bar\Delta} = \coth \bar \Delta/2T_r$. Solid line corresponds to
Eq.~\eqref{admittance_rate} with non-equilibrium $B_{\bar \Delta}$
given by Eq.~\eqref{e-h-distr}. We use $g=0.5$ and $\Omega= 0.02
E_c$. See text.
            }\label{figureDelsing}
\end{figure}

The dissipative part of the admittance in a SEB has been addressed
experimentally via radio-frequency reflectometry measurements. The
device was exposed to a continuous rf-signal~\cite{delsing}. In the
experiment the tunneling conductance was estimated to be equal
$g=0.5$ such that the SEB was in the strong Coulomb blockade regime.
We plot the real part of the admittance~\eqref{admittance_rate} at
fixed $\Omega$ as a function of $q$ in Fig.~\ref{figureDelsing}.
There, for a sake of comparison, we present $\Re \mathcal{G}_\Omega$
computed i) in the equilibrium without taking into account the
renormalization effects, i.e., with $Z=1$ and $B_{\bar\Delta} =
\coth \Delta/2T_r$ (dashed line); ii) in the equilibrium and with
the renormalization effects, i.e., with $B_{\bar\Delta} = \coth
\bar\Delta/2T_r$ (dotted line); iii) out of equilibrium with
$T_d>T_r$ and $B_{\bar\Delta}$ determined by Eq.~\eqref{e-h-distr}
(solid line). In all three cases, we use the same values of $g$,
$E_c$ and $\Omega$ corresponding to the experiment~\cite{delsing}.
As one can see from Fig.~\ref{figureDelsing}, a slight variation of
ratios $C_g/C$ and $T_{r,d}/E_c$ allows us to make curves for cases
i), ii) and iii) indistinguishable. In Ref.~\cite{delsing} it is
assumed that the electrons inside the island are in the equilibrium
with the reservoir and the renormalization effects are not important
(case i) above). Values of $C_g/C$ and $T_r/E_c$ are used as fitting
parameters. The curves presented in Fig.~\ref{figureDelsing} however
demonstrate a more subtle picture. As one can see, the successful
fitting of the experimental data by a `theoretical' curve gives yet
no confidence that these assumptions are satisfied. Therefore, more
careful analysis of the experimental data of Ref.~\cite{delsing} is
needed.

%%%%%%%%%%%%%%%%%%%%%
%%%%%%%%%%%%%%%%%%%

The electron-hole distribution $B_\omega$ enters admittance in a
twofold way. The analytical structure of admittance as a function of
external frequency $\Omega$ is entirely determined by $B_\omega$ at
$\omega=-{\bar\Delta}$. The scaling parameter $Z$ arising from the
renormalization however contains information on $B_\omega$ in wide
domain $\omega_0<|\omega| < E_c$. Admittance~\eqref{admittance_rate}
can serve as a tool for direct experimental measurement of
$B_\omega$. As such can be the measurement of a real part of
admittance $\Re \mathcal{G}_\Omega$ at two different driving
frequencies. Other possibility would be the simultaneous measurement
of the real and imaginary parts of $\mathcal{G}_\Omega$ at a given
frequency~\cite{gabelli}. Then one can read out $\bar\Delta
B_{-\bar\Delta}$ in the entire span of $\bar\Delta$ by tuning the DC
gate voltage $U_0$. Measurements of frequency dependence of $\Re
\mathcal{G}_\Omega$ at the Coulomb peak ($\bar\Delta=0$) provide an
access to the effective bosonic temperature $T_{\rm eh}$. Thus the
admittance of a SEB under AC gate voltage can be used as the
thermometer for the electron-hole excitations similar to the Coulomb
blockade thermometer based on measurements of the differential DC
conductance in a single electron transistor
(SET)~\cite{Giazotto_review}.

%\emph{Dissipation, Charge relaxation resistance, renormalized capacitance.---}

The real part of admittance determines energy dissipation rate:
$\mathcal{W}_\Omega=(C_g/2C)\Re{\cal G}_\Omega |U_\Omega|^2$. At
quasi-static regime $ \Omega\to 0$ we find that even out of
equilibrium the energy dissipation rate factorizes into the product
of well-defined physical observables in full analogy with classical
expression~\eqref{admittance0}:
\begin{equation}
  \label{distwo}
  \mathcal{W}_\Omega=\frac{\Omega^2}{2}
  \mathcal{C}^2_g\mathcal{R}_q |U_\Omega|^2,\quad
\mathcal{R}_q=\frac{h}{e^2 g^\prime}, \quad
 \mathcal{C}_g = \frac{\partial q^\prime}{\partial U_0}.
\end{equation}
The charge relaxation resistance $\mathcal{R}_q$ and the renormalized gate
capacitance $\mathcal{C}_g$ are related to physical observables formally
defined as~\cite{burmistrov1}
\begin{equation}
  \label{quantities}
    g^\prime=4\pi\Im\frac{\partial K^R(\omega)}{\partial\omega}
    %\Big|_{\omega=0}
     ,\quad
    q^\prime=Q+\Re\frac{\partial
    K^R(\omega)}{\partial\omega} .
    %\Big|_{\omega=0}.
\end{equation}
Here $Q = \langle\hat{n}_d\rangle$ is the average charge in the
island, the correlation function $
K^R(t)=i\theta(t)\langle[X(t),X^\dag(0)]\rangle$ and the limit
$\omega\to 0$ is assumed. The physics behind
quantities~\eqref{quantities} can be understood if we turn from a
SEB to a SET. In the absence of source-drain voltage a SET is
physically equivalent to the SEB. The quantity\ $g^\prime$\ then
coincides with the SET conductance~\cite{schon1,schon2}. The
quantity\ $q^\prime$ is specific to Coulomb blockade physics and can
be addressed as the quasi-particle charge~\cite{burmistrov1}.

With the help of definitions~\eqref{quantities} we obtained the
following results in the out-of-equilibrium regime (for $g\ll 1$)
\begin{equation}
  g^\prime= -\frac{1}{2} \bar g\bar\Delta \partial_{\bar\Delta} \ln
  B_{-\bar\Delta},\qquad
  q^\prime=k+\frac{1}{2}+\frac{1}{2}\frac{1}{B_{-\bar{\Delta}}}. \label{gPrqPr}
\end{equation}
Equations~\eqref{gPrqPr} generalize the results for
$g^\prime$~\cite{schon1,schon2} and
$q^\prime$~\cite{burmistrov1} derived under the
equilibrium conditions.

\emph{Derivation.} -- Below we describe the main steps of the derivation. Further details
will be given in~\cite{tbp}. Following Ref.~\cite{matveev}, we
write the Hamiltonian \eqref{ham1} in the truncated Hilbert space of
electrons on the island accounting for two charging states: with\
$Q=k$\ and\ $Q=k+1$ only(see Fig.~\ref{figure2}). The projected
Hamiltonian then takes a form of $2\times2$\ matrix acting in the
space of these two charging states:
\begin{gather}
  \label{ham4}
   {\cal H}=H_0 + H_t + \Delta S_z + \Delta^2/4E_c
\end{gather}
where
\begin{equation}
  \label{ham5}
 {\cal H}_t=\sum_{k,\alpha}t_{k\alpha} a^\dagger_k d_\alpha
   S^-+\hbox{H.c.}
\end{equation}
and $S^z,\ S^{\pm}=S^x\pm iS^y$\ are ordinary (iso)spin\ $1/2$\
operators. Admittance is proportional to dynamical (iso)spin
susceptibility $\Pi_s^R(t)=i\theta(t)\langle
[S^z(t),S^z(0)]\rangle$~\cite{rodbur}:
\begin{equation}
 \mathcal{G}_\Omega=-i\Omega C_g\Pi_{s}^{R}(\Omega)/C.
\label{spin-dissipation}
\end{equation}
To deal with spin operators out of equilibrium the generalization of
Abrikosov's pseudo-fermions (PF) $\psi^\dagger_\alpha$, $\psi_\alpha$ is
used~\cite{wingreen,woelfle}. Integrating out electrons in the limit
$N_{\rm ch}\gg1$, we arrive at the following effective
action~\cite{rodbur2}
\begin{gather}
    S=
  \int dt\bar{\psi}\Big(i\partial_t-\frac{\sigma_z \Delta}{2}+\eta\Big)\psi +\frac{g}{8}\int\bar{\psi}(t)\gamma_i\sigma_-\psi(t)
  \notag\\
  \times\Pi_{ij}(t,t^\prime)
   \bar{\psi}(t^\prime)\gamma_j\sigma_+\psi(t^\prime)\,dtdt^\prime.
 \label{pseudo-action1}
\end{gather}
Here the pseudo-fermion fields $\psi$,\ $\bar{\psi}$ are understood
as vectors in the tensor product of isospin and Keldysh spaces. We
inserted the factor\ $\exp(\eta \bar{\psi}\psi)$\ with
$\eta\to-\infty$ into the density matrix in order to fulfill the
constraint $\bar{\psi}(t)\psi(t)=1$. The matrices $\sigma_z$,\
$\sigma_\pm=(\sigma_x\pm i\sigma_y)/2$, and $\gamma_1\equiv\tau_x,\
\ \gamma_2\equiv\tau_0$ are the Pauli matrices in (iso)spin and
Keldysh spaces respectively. $\Pi_{ij}$ stands for the matrix:
\begin{gather}
  \label{aes2}
%    \begin{split}
  \Pi=\begin{pmatrix}
                  0     & \Pi^A\\
                  \Pi^R & \Pi^K
                \end{pmatrix},\\
   \Pi^{R,A,K}(t,t^\prime)=\int\frac{d\omega}{2\pi}\Pi^{R,A,K}_\omega(\tau)
   e^{-i\omega(t-t^\prime)} ,\\
   \Pi^{R,A}_\omega(\tau)= \mp i\int
   \big[F^d_\varepsilon(\tau)-F^r_{\varepsilon-\omega}(\tau)\big]
   \frac{d\varepsilon}{2\pi},\label{aes2a} \\
        \Pi^K_\omega(\tau)= 2i\int(1-F^d_\varepsilon(\tau)
        F^r_{\varepsilon-\omega}(\tau))\frac{d\varepsilon}{2\pi}.\label{aes2b}
%  \end{split}
\end{gather}
The PF dynamical spin susceptibility is given
as~\cite{tbp}:
\begin{gather}
   \Pi^R_{s,pf}(\omega)=Z^2\sum_\sigma\int\Biggr \{
   \bm{\Gamma}^{RKR}_\sigma(
   \varepsilon+\omega,\varepsilon,\omega)
   \overline{G}^R_{\sigma,\varepsilon+\omega}\overline{G}^R_{\sigma,\varepsilon}\notag\\
   +\bm{\Gamma}^{RAR}_\sigma(\varepsilon+\omega,\varepsilon,\omega)
   \Bigl [ \overline{G}^R_{\sigma,\varepsilon+\omega}\overline{G}^K_{\sigma,\varepsilon}+
   \overline{G}^K_{\sigma,\varepsilon+\omega}\overline{G}^A_{\sigma,\varepsilon}\Bigr ]\notag\\
   +
   \bm{\Gamma}^{KAR}_\sigma(\varepsilon+\omega,\varepsilon,\omega)\overline{G}^A_{\sigma,\varepsilon+\omega}\overline{G}^A_{\sigma,\varepsilon}\Biggl \}\frac{d\varepsilon}{16\pi i} ,
   \label{polar1}
\end{gather}
where the renormalized
Green's
function~\cite{rodbur}
\begin{gather}
   \overline{G}_{\sigma,\varepsilon}^{R,A}=\frac{Z(\lambda)}{\varepsilon-\bar{\xi}_\sigma\pm i\bar{g}
   \Gamma_\sigma(\varepsilon)}, \qquad
   \bar{\xi}_\sigma=-\eta+\sigma\bar{\Delta}/2,\notag\\
  \Gamma_\sigma(\varepsilon)=
   \frac{1}{8\pi}(\varepsilon-\bar\xi_{-\sigma})
   [{\cal F}^{-\sigma}_{\bar{\xi}_{-\sigma}}+B_{\varepsilon-\bar{\xi}_{-\sigma}}] .
\end{gather}
The pseudo-fermion distribution\ ${\cal F}^\sigma_{\varepsilon}$ is
not known a priori.  It is to be determined self-consistently from
corresponding kinetic equation. It obeys~\cite{rodbur2}:
\begin{gather}
  \label{self-con0}
  \begin{split}
   {\cal F}^\sigma_{\varepsilon}=\frac{B_{-\sigma(\varepsilon+\frac{\Delta\sigma}{2}+\eta)}
   {\cal F}^{-\sigma}_{\bar{\xi}_{-\sigma}}-\sigma}
   {B_{-\sigma(\varepsilon+\frac{\Delta\sigma}{2}+\eta)}-\sigma
  {\cal F}^{-\sigma}_{\bar{\xi}_{-\sigma}}}.
  \end{split}
\end{gather}
As was shown in~\cite{rodbur} all terms of $G^RG^R$\ and $G^AG^A$
type are controlled by renormalization scheme and can be discarded.
Then, Eq.~\eqref{polar1} becomes simplified:
\begin{equation}
  \label{polar2}
   \Pi^R_{s,pf}(\omega)=\frac{Z^4}{8}\sum_\sigma\partial_{\bar{\xi}_{\sigma}}
   {\cal F}^{\sigma}_{\bar{\xi}_{\sigma}}
   \Big[1-\frac{\omega\bm{\Gamma}^{RAR}_\sigma(\bar{\xi}_\sigma+\omega,\bar{\xi}_\sigma,\omega)}
   {\omega+2i\bar{g}\Gamma_\sigma}\Big]
\end{equation}
where $\Gamma_\sigma=\Gamma_\sigma(\bar{\xi}_\sigma)$.
The vertex function $\bm{\Gamma}^{RAR}$ solves  the following Dyson equation
\begin{gather}
    \bm{\Gamma}^{RAR}_\sigma(\varepsilon+\omega,\varepsilon,\omega)=1+\frac{ig}{4}\int\frac{dx}{2\pi}
       \overline{G}^R_{-\sigma,\varepsilon+\omega+x}\overline{G}^A_{-\sigma,\varepsilon+x} \notag\\
     \times
%     \Big[\Pi_x^A\mathcal{F}^{-\sigma}_{\varepsilon+\omega+x}-\Pi_x^R\mathcal{F}^{-\sigma}_{\varepsilon+x}+\Pi^K_x\Big]
\Im \Pi_x^R (B_x-\sigma)  \bm{\Gamma}^{RAR}_{-\sigma}(\varepsilon+\omega+x,\varepsilon+x,\omega) .
     \label{vertex1}
\end{gather}
By using Eqs~\eqref{self-con0}-\eqref{polar2} and the solution of Eq.~\eqref{vertex1}:
\begin{equation}
  \label{vertex-solution1}
\frac{\bm{\Gamma}^{RAR}_\sigma(\bar\xi_\sigma+\omega,\bar\xi_\sigma,\omega)}{\omega+2i\bar{g}\Gamma_\sigma}=
\frac{1}{\omega}
   \frac{\omega+2i\bar{g}(\Gamma_{-\sigma}-\Gamma_{\sigma})}{\omega+2i\bar{g}(\Gamma_{-\sigma}+\Gamma_{\sigma})} ,
\end{equation}
we obtain expression~\eqref{admittance_rate} for the admittance.

The computation of $q^\prime$ and $g^\prime$ is entangled with the
computation of $K^R_{\omega}$ (see Eq.~\eqref{quantities}). Using
the definition of $K^R(t)$ in terms of the operators  $X(t)$, one
can obtain the following expression~\cite{tbp}:
\begin{gather}
   K^R_{\omega}=-\frac{g}{8\pi}
      \int\frac{d\omega^\prime}{2\pi} \Bigl [
      i\Im {\cal
      D}_{\omega^\prime}(B_{\omega^\prime}-B_{\omega^\prime-\omega})
\notag \\
+  \Re{\cal
      D}_{\omega^\prime}B_{\omega^\prime-\omega} \Bigr ]
      \int(F^d_{\varepsilon+\omega^\prime-\omega}-F^{r}_\varepsilon)d\varepsilon .
 \label{kr1}
\end{gather}
Here we introduce the transverse spin susceptibility ${\cal
D}^R(t)=i\theta(t)\langle[S^-(t),S^+(0)]\rangle$.
Following Eq.~\eqref{quantities} one straight forwardly
establishes:
\begin{gather}
  g^\prime=g
      \int\frac{d\omega}{2\pi}\hbox{Im}{\cal D}^R_\omega\omega\partial_\omega B_\omega,
      \label{conduct0}\\
  q^\prime=Q+\frac{g}{4\pi}
      \int\frac{d\omega}{2\pi}\hbox{Re}{\cal
      D}^R_{\omega}\partial_{\omega}(\omega
      B_{\omega}).\label{qprime0}
\end{gather}
The average charge in the island is given in terms of the average isospin as:
%\begin{gather}
  $Q=k+1/2-\langle S_z\rangle$.
%\end{gather}
Using the result for the transverse spin susceptibility
~\cite{rodbur2}:
\begin{gather}
  \label{dr}
  {\cal D}^R_\omega=\frac{1}{B_{-\bar\Delta}}\frac{Z^2(\lambda)}{\omega+\bar{\Delta}+i0^+} ,
\end{gather}
we obtain results~\eqref{gPrqPr} for $g^\prime$ and $q^\prime$.

In summary, the paper addresses  the admittance and energy
dissipation in an out-of-equlibrium single electron box under strong
Coulomb blockade ($g\ll 1$). We deal with the regime when electron
coherence can be neglected but quantum fluctuations of charge are
strong due to Coulomb interaction. We derived the expression for the
admittance at frequencies $\Omega \ll
\max\{T_r,\varepsilon_d,|\bar\Delta|\}$. We found that the energy
dissipation rate retains its universal appearance in the
quasi-stationary limit even out of equilibrium. It is achieved in
terms of specially chosen physical observables: the charge
relaxation resistance and the renormalized gate capacitance. We
propose the admittance as a tool for a measurement of the effective
bosonic distribution corresponding to electron-hole excitations in
the system.

The authors are grateful to A. Ioselevich, Yu. Makhlin, and J.
Pekola for stimulating discussions. The research was funded in part
by the Russian Ministry of Education and Science under Contract No.
P926, the Council for Grant of the President of Russian Federation
Grant No. MK-125.2009.2, RFBR Grants No. 09-02-92474-MHKC and RAS
Programs ``Quantum Physics of Condensed Matter'' and ``Fundamentals
of nanotechnology and nanomaterials''.

\end{document}